\documentclass[11pt]{article}
\usepackage{moriond,amsmath,amssymb,bbm}

\newcommand{\tmop}[1]{\ensuremath{\operatorname{#1}}}
\newcommand{\tmem}[1]{{\em #1\/}}
\newcommand{\mathpi}{\pi}
\newcommand{\mathi}{\mathrm{i}}
\newcommand{\mathe}{\mathrm{e}}
\newcommand{\emdash}{---}

\begin{document}

\hfill
\texttt{IFIC/05-35}

\hfill
\texttt{FTUV/05-0719}\\

\vspace*{2.5cm}
\title{Higgs-less Higgs mechanism: low-energy expansion
\footnote{ Talk given by JH at 40$^\text{th}$ {\it{Rencontres de Moriond}}, Electroweak Session, March 5-12 2005,  La Thuile, Aosta Valley, Italy}
}

\author{Johannes Hirn$^\S$ and Jan Stern$^\P$}

\address{$^\S$IFIC, Departament de F\'\i sica Te\`orica, CSIC - Universitat de Val\`encia\\ Edifici d'Instituts de Paterna, Apt. Correus 22085, 46071 Val\`encia, Spain\\
$^\P$Groupe Physique Th\'eorique, Unit\'e mixte de recherche 8608 du CNRS\\ IPN Orsay, Universit\'e~Paris-Sud~XI, 91406~Orsay, France}

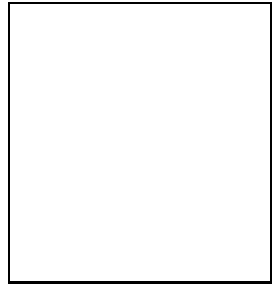
\maketitle\abstracts{In this talk, we describe an effective theory for electroweak symmetry breaking without a physical Higgs, based on a symmetry larger than the electroweak gauge group. This symmetry forbids deviations from the Standard Model at the leading order in the appropriate chiral expansion. Indeed, the large symmetry allows for a consistent expansion  of the effective theory in powers of momenta and spurions. The latter are automatically present: they define the covariant reduction from the large symmetry to the electroweak group.}

\section{Introduction}

We consider electroweak symmetry breaking (EWSB) without a Higgs boson: the
Higgs mechanism removes the three Goldstone bosons (GBs) from the
spectrum, while the usual Higgs boson does not exist. We wish to construct a
low-energy effective theory (LEET), expanding in powers of momenta (and other
naturally-small parameters, the spurions, as we shall soon see). Indeed, for
the expansion to be consistent according to the rules of a LEET (such as
Chiral Perturbation Theory {\cite{Weinberg:1979kz,Gasser:1984yg}}), we need to
assume the presence of a symmetry $S_{\text{nat}}$ controlling the smallness
of deviations from the Standard Model (SM), i.e. technical naturalness.

We require the ``hidden'' symmetry $S_{\text{nat}}$ to be sufficiently large
as to force the leading $\mathcal{O} \left( p^2 \right)$ order of the LEET to
coincide with the tree-level Higgs-less vertices of the SM in the limit of
vanishing fermion masses. The $S_{\text{nat}}$ symmetry can be viewed as
defining the custodial symmetry (and its extension to the left-handed
non-abelian sector) in the presence of non-vanishing gauge coupling $g'$ ($g$).
$S_{\text{nat}}$ necessarily contains the SM gauge group $\tmop{SU} \left( 2
\right)_L \times \mathrm{U} \left( 1 \right)_Y$, which we call
$S_{\text{red}}$. The higher symmetry $S_{\text{nat}} \supset S_{\text{red}}$
can be linearized by adding a set of nine auxiliary gauge fields to the
original four present in the SM. The additional gauge fields are not physical:
they will be eliminated by constraints, implemented via spurions.

Once the constraints are applied, the theory only contains $\tmop{SU} \left( 2
\right)_L \times \mathrm{U} \left( 1 \right)_Y$ Yang-Mills fields and chiral
fermions coupled to three Goldstone bosons $\Sigma(x)$. The latter disappear from the spectrum,
resulting in three of the vector fields acquiring a mass. All physical degrees
of freedom are light compared to the scale $\Lambda_{\text{w}} \simeq 4
\mathpi v \simeq 3$ TeV: the only new particles beyond those already known are
{\tmem{light}} right-handed neutrinos.

This talk is based on the first half of {\cite{Hirn:2005fr}}. In that paper,
we also studied in detail some of the consequences for lepton-number violation
(LNV) processes, vertex corrections (which come in before oblique corrections
in the present case) as well as the possible contribution of light
right-handed neutrinos to dark matter.

\section{Higgs-less LEET based only on $\tmop{SU} \left( 2 \right)_L \times
\mathrm{U} \left( 1 \right)_Y$}

\subsection{Power-counting} \label{pow-count}

In the case of Higgs-less EWSB, we do not have a renormalizable theory to
start with: all operators respecting the symmetries must be included in the
effective lagrangian. The infinite number of them should be ordered according
to their importance in the low-energy limit $p \rightarrow 0$, i.e. according
to their infrared (or chiral) dimension $d_{\text{IR}}$. The effective
lagrangian is then expressed as $\mathcal{L}_{\text{eff}} =
\sum_{d_{\text{IR}} \geqslant 2} \mathcal{L}_{d_{\text{IR}}}$ with
$\mathcal{L}_{d_{\tmop{IR}}} =\mathcal{O} \left( p^{d_{\tmop{IR}}} \right)$. A
local operator/interaction vertex $\mathcal{O}$ built from GBs, gauge fields
and fermions, carries the infrared dimension
\begin{eqnarray}
  d_{\text{IR}} \left[ \mathcal{O} \right] & = & n_{\partial} \left[
  \mathcal{O} \right] + n_g \left[ \mathcal{O} \right] + \frac{1}{2} n_f
  \left[ \mathcal{O} \right],  \label{3.16a}
\end{eqnarray}
where $n_{\partial} \left[ \mathcal{O} \right]$ is the number of derivatives
entering the operator $\mathcal{O}$, $n_g \left[ \mathcal{O} \right]$ the
number of gauge coupling constants and $n_f \left[ \mathcal{O} \right]$ the
number of fermion fields. This infrared counting rule provides the basis for
the ordering of diagrams. Provided fermion
masses can consistently be counted as $\mathcal{O} \left( p^1 \right)$ or
higher (more on this later), one finds that the degree of suppression
$d_{\tmop{IR}} [ \Gamma ]$ of the diagram $\Gamma$ is given by
{\cite{Wudka:1994ny}}
\begin{eqnarray}
  d_{\text{IR}} \left[ \Gamma \right] & = & 2 + 2 L + \sum^V_{v = 1} \left(
  d_{\text{IR}} \left[ \mathcal{O}_v \right] - 2 \right),  \label{3.28}
\end{eqnarray}
where $L$ is the number of loops and $d_{\text{IR}} \left[ \mathcal{O}_v
\right]$ is the infrared dimension (\ref{3.16a}) of the vertex
$\mathcal{O}_v$, and the vertices are numbered $v = 1, \cdots, V$. The
expansion in powers of $d_{\tmop{IR}}$ is also an expansion in loops $L$: it
makes sense at least at the formal level provided {\tmem{all}} operators
invariant under the symmetry have $d_{\tmop{IR}} \left[ \mathcal{O} \right]
\geqslant 2$.

\subsection{Unwanted operators at leading order} \label{unwanted}

Among all $\tmop{SU} \left( 2 \right)_L \times \mathrm{U} \left( 1
\right)_Y$-invariant operators of lowest order in the Higgs-less theory, one
finds (at leading order) operators which have no equivalent in the
renormalizable lagrangian of the SM. In fact, the corresponding operators can
be built using SM fields, but they would have mass-dimension six.
In the absence of the Higgs particle, this suppression no longer holds.

Using a left-handed lepton doublet $\ell_L$, one can translate Weinberg's
{\cite{Weinberg:1979sa}} LNV $\tmop{SU} \left( 2 \right)_L \times \mathrm{U}
\left( 1 \right)_Y$-invariant operator to the Higgs-less case, obtaining
$\Lambda \overline{\ell_L} \Sigma \tau^+ \Sigma^{\dag}  \left( \ell_L
\right)^c =\mathcal{O} \left( p^1 \right)$. According to the power-counting
rules given above, this operator appears at $\mathcal{O} \left( p^1 \right)$,
without any suppression factor. Other operators that can be written at
$\mathcal{O} \left( p^1 \right)$ yield Dirac masses to fermions $\Lambda
\overline{\chi_L} \Sigma \left( \tau^3 \right) \chi_R$. Such operators have
chiral dimension less than two, and therefore endanger the internal
consistency of the expansion procedure, as mentioned in Section \ref{pow-count}.

At $\mathcal{O} \left( p^2 \right)$, one finds other ``unwanted'' operators
involving fermions. Non-universal couplings~{\cite{Appelquist:1985rr,Peccei:1990kr}} to massive vector bosons appear at
$\mathcal{O} \left( p^2 \right)$: $\mathi \overline{\chi_L} \gamma^{\mu} 
\left( \Sigma D_{\mu} \Sigma^{\dag} \right) \chi_L$ and $\mathi
\overline{\chi_R} \gamma^{\mu}  \left( \Sigma^{\dag} D_{\mu} \Sigma \right)
\chi_R$. In addition, this introduces couplings of the right-handed fermions
to the $W^{\pm}$. Both types of operators would also be a new source of
flavor-changing currents, requiring a redefinition of the CKM matrix, which
would not be unitary anymore. These operators would also introduce
flavor-changing neutral currents (FCNCs) at this level.

At $\mathcal{O} \left( p^2 \right)$, we find two more operators, giving
tree-level contributions to the $S$ {\cite{Holdom:1990tc,Peskin:1992sw}} and
$T$ {\cite{Longhitano:1980iz}} parameters. These operators are $\tmop{SU} \left( 2
\right)_L \times \mathrm{U} \left( 1 \right)_Y$-invariant, but break custodial symmetry.

\section{Higgs-less LEET based on $S_{\text{nat}}$} \label{Snat}

\subsection{The symmetry $S_{\text{nat}}$}

We assume a larger hidden symmetry $S_{\text{nat}} \supset \tmop{SU} \left( 2
\right)_L \times \mathrm{U} \left( 1 \right)_Y$: this allows for an expansion procedure consistent with the
principles of a LEET, and in which the unwanted operators of the previous
Section are relegated to higher orders. In the minimal version, and before the
constraints are applied, the lagrangian of the theory at $\mathcal{O} \left(
p^2 \right)$ consists of two decoupled sectors: a) the symmetry-breaking
sector containing three GBs together with six connections of the
spontaneously-broken $\tmop{SU} \left( 2 \right)_{\Gamma_L} \times \tmop{SU}
\left( 2 \right)_{\Gamma_R}$ symmetry {\tmem{and}} b) an unbroken $\tmop{SU}
\left( 2 \right)_{G_L} \times \tmop{SU} \left( 2 \right)_{G_R} \times
\mathrm{U} \left( 1 \right)_{B - L}$ gauge theory with the $L \leftrightarrow
R$ symmetric coupling of local left and right isospin to chiral fermion
doublets. The symmetry group $S_{\text{nat}}$ is thus
\begin{eqnarray}
  S_{\text{nat}} & = & \left[ \tmop{SU} \left( 2 \right) \times \tmop{SU}
  \left( 2 \right) \right]^2 \times \mathrm{U} \left( 1 \right)_{B - L} . 
  \label{1.6}
\end{eqnarray}
We assume an underlying theory responsible for the spontaneous symmetry
breaking of $\tmop{SU} \left( 2 \right)_{\Gamma_L} \times \tmop{SU} \left( 2
\right)_{\Gamma_R} \subset S_{\text{nat}}$ down to its vector subgroup. This
produces a triplet of GBs $\Sigma$ transforming according to {\footnote{The
notation $\Sigma$ is the same as in the previous Section, but the
transformation properties are different.}}
\begin{eqnarray}
  \Sigma & \longmapsto & \Gamma_L \Sigma \Gamma_R^{\dag},  \label{00000}
\end{eqnarray}
where $\Gamma_L \in \tmop{SU} \left( 2 \right)_{\Gamma_L}$ and $\Gamma_R \in
\tmop{SU} \left( 2 \right)_{\Gamma_R}$, and the corresponding connections are
denoted by $\Gamma_{L \mu}, \Gamma_{R \mu}$. So much for the composite sector
of the theory.

On the other hand, for the elementary sector of the theory, the elementary
fermion doublets $\chi_{L, R}$ transform as
\begin{eqnarray}
  \chi_{L, R} & \longmapsto & G_{L, R} \mathe^{- \mathi \frac{B - L}{2}
  \alpha} \chi_{L, R} .  \label{cp}
\end{eqnarray}
The gauge fields of the elementary group $\tmop{SU} \left( 2 \right)_{G_L}
\times \tmop{SU} \left( 2 \right)_{G_R} \times \mathrm{U} \left( 1 \right)_{B
- L}$ are denoted by $g_L G_{L \mu}, g_R G_{R \mu}, g_B G_{B \mu}$, and we
thus collect all $\mathcal{O} \left( p^2 \right)$ terms as
\begin{eqnarray}
  \mathcal{L} \left( p^2 \right) & = & \frac{f^2}{4}  \left\langle D_{\mu}
  \Sigma^{\dag} D^{\mu} \Sigma \right\rangle + \mathi \overline{\chi_L}
  \gamma^{\mu} D_{\mu} \chi_L + \mathi \overline{\chi_R} \gamma^{\mu} D_{\mu}
  \chi_R \nonumber\\
  & - & \frac{1}{2}  \left\langle G_{L \mu \nu} G_L^{\mu \nu} + G_{R \mu \nu}
  G^{\mu \nu}_R \right\rangle - \frac{1}{4} G_{B \mu \nu} G_B^{\mu \nu} . 
  \label{lag}
\end{eqnarray}

We see that the symmetry $S_{\text{nat}}$ eliminates all the unwanted
couplings discussed in Section~\ref{unwanted} at the leading chiral order
$\mathcal{O} \left( p^2 \right)$ described by the lagrangian (\ref{lag}). On
the other hand, it seems at first sight that $S_{\text{nat}}$ is too large:
the lagrangian (\ref{lag}) contains thirteen gauge connections $g_L G_{L \mu},
g_R G_{R \mu}, g_B G_{B \mu}, \Gamma_{L \mu}, \Gamma_{R \mu}$, as compared to
four in the SM. Due to the GB term in (\ref{lag}) (first term in the
right-hand side), the three combinations $\Gamma_{R \mu}^a - \Gamma_{L \mu}^a$
acquire a mass term by the Higgs mechanism, whereas all ten remaining vector
fields as well as fermions remain massless. Furthermore, the lagrangian
(\ref{lag}) does not contain any coupling that would transmit the symmetry
breaking from the composite sector ($\Sigma, \Gamma_{L \mu}, \Gamma_{R \mu}$)
to the elementary sector ($G_{L \mu}, G_{R \mu}, G_{B \mu}, \chi$). We now
remedy this by introducing constraints that reduce the space of gauge
connections.

\subsection{Reduction $S_{\text{nat}} \rightarrow S_{\text{red}}$ via
constraints}

We want to identify $\Gamma_{L \mu}$ to $g_L G_{L \mu}$, up to a gauge
transformation $\Omega_L \in \tmop{SU} \left( 2 \right)$, i.e.
\begin{eqnarray}
  \Gamma_{L \mu} & = & \Omega_L \left( x \right) g_L G_{L \mu} \Omega_L^{- 1}
  \left( x \right) + \mathi \Omega_L \left( x \right) \partial_{\mu}
  \Omega_L^{- 1} \left( x \right) .  \label{39}
\end{eqnarray}
This will reduce the group $\tmop{SU} \left( 2 \right)_{G_L} \times \tmop{SU}
\left( 2 \right)_{\Gamma_L}$ to its vector subgroup, which will be recognized
as the $\tmop{SU} \left( 2 \right)_L$ of the SM. Requiring the invariance of
the constraint (\ref{39}) with respect to the whole symmetry $\tmop{SU} \left(
2 \right)_{G_L} \times \tmop{SU} \left( 2 \right)_{\Gamma_L}$ amounts to
promoting the gauge function $\Omega_L$ to a field that transforms according
to
\begin{eqnarray}
  \Omega_L & \longmapsto & \Gamma_L \Omega_L G_L^{\dag} .  \label{40}
\end{eqnarray}
The field $\Omega_L$ is {\tmem{not}} a GB, but rather a non-propagating
{\tmem{spurion}}: this follows from the constraint~(\ref{39}) itself, since
the latter can be equivalently rewritten as
\begin{eqnarray}
  D_{\mu} \Omega_L & \equiv & \partial_{\mu} \Omega_L - \mathi \Gamma_{L \mu}
  \Omega_L + \mathi g_L \Omega_L G_{L \mu} \hspace{1em} = \hspace{1em} 0 . 
  \label{41}
\end{eqnarray}
The spurion has no dynamics, since no kinetic term can be written down for it.
A similar procedure can be performed in the right-handed sector: it is
complicated by the selection of the $\mathrm{U} \left( 1 \right)$ subgroups
{\cite{Hirn:2005fr}}, so we do not describe it here.

The problem of unwanted terms reappears as long as the spurion $\mathcal{X}$
is restricted to be unitary: one can still construct other
$S_{\text{nat}}$-invariants that are $\mathcal{O} \left( p^2 \right)$ but are
not contained in~(\ref{lag}). These additional terms are exactly all the
unwanted terms of Section \ref{unwanted}, which still have to be suppressed.
This can be achieved by adding a new ingredient: we now admit multiplication
of the unitary spurions by constants which are (technically) naturally small.
This is implemented via the requirement that only the object $\mathcal{X}$
\begin{eqnarray}
  \mathcal{X} & \equiv & \xi \Omega_L,  \label{X}
\end{eqnarray}
may be inserted in the operators of Section \ref{unwanted} in order to make
them invariant under $S_{\text{nat}}$. The order of magnitude of $\xi$ should
later be estimated from experiments, but $\xi$ will be considered as an expansion
parameter. We now require, by analogy with (\ref{41})
\begin{eqnarray}
  D_{\mu} \mathcal{X} & \equiv & \partial_{\mu} \mathcal{X}- \mathi \Gamma_{L
  \mu} \mathcal{X}+ \mathi g_L \mathcal{X}G_{L \mu} \hspace{1em} =
  \hspace{1em} 0 .  \label{c1}
\end{eqnarray}
This implies the existence of a ``standard gauge'', specified by $\Omega_L
= 1$, in which the connections are equal, i.e. $\Gamma_{L \mu}
\overset{\text{s.g.}}{=} g_L G_{L \mu}$: in this gauge the spurion $\mathcal{X}$ reduces to one constant $\xi$.

For the right-handed sector, one has to define two spurions $\mathcal{Y}$ and
$\mathcal{Z}$. One may again define the standard gauge, in which the spurions
reduce to two constants $\eta$ and $\zeta$ respectively, while the gauge
connections are identified appropriately according to $\Gamma^{1, 2}_{R \mu}
\overset{\text{s.g.}}{=} g_R G_{R \mu}^{1, 2} \overset{\text{s.g.}}{=} 0
\text{}$ and $\Gamma^3_{R \mu} \overset{\text{s.g.}}{=} g_R G_{R \mu}^3
\overset{\text{s.g.}}{=} g_B G^3_{B \mu}$.

\section{Consequences of the formalism}

\subsection{Construction of the LEET}

The next step in the formulation of the LEET is the construction of the
effective lagrangian: one writes down all terms invariant under
$S_{\text{nat}}$ that can be constructed out of the GBs $\Sigma$, the
connections $\Gamma_{L \mu}, \Gamma_{R \mu}$, the gauge fields $G_{L \mu},
G_{R \mu}, G_{B \mu}$, the spurions $\mathcal{X},\mathcal{Y},\mathcal{Z}$, and
fermions. The operators should be ordered according to their chiral
power-counting and to the powers of spurions involved. To exhibit the physical
content of each operator, one then injects the solution of the constraints in
the standard gauge. This yields a lagrangian depending on the fermions and on
the $S_{\text{red}}$ gauge fields, which should be used as dynamical variables
to compute loops. In addition, this lagrangian depends on the three constants
$\xi$, $\eta$ and $\zeta$. At the leading order $\mathcal{O} \left( p^2
\right)$ without explicit powers of spurions, the lagrangian describes exactly
the SM couplings but without the Higgs boson, and with all fermions left
massless. The origin of fermion masses, which will come with explicit powers
of spurions, thus appears different from that of vector bosons. Other terms
involving explicit powers of spurions will also bring other interactions
{\cite{Hirn:2005fr}}.

\subsection{Dirac masses}

To be invariant under $S_{\text{nat}}$, Dirac masses require one insertion of
the spurion $\mathcal{X}$ and one of the spurion $\mathcal{Y}$  ($\mathcal{Y}$
and $\mathcal{Y}_c \equiv \tau^2 \mathcal{Y}^{\ast} \tau^2$ can be
conveniently decomposed as linear combinations of $\mathcal{Y}_u$ and
$\mathcal{Y}_d$). Hence, the leading quark mass term in the lagrangian is of
order $\mathcal{O} \left( p^1 \xi^1 \eta^1 \right)$ and reads
\begin{eqnarray}
  \mathcal{L} \left( p^1 \xi^1 \eta^1 \right)_{\text{quarks}} & = & -
  \Lambda_{\text{quarks}}  \left( \overline{q_L} \mathcal{X}^{\dag} \Sigma
  \mathcal{Y}_u q_R + \overline{q_L} \mathcal{X}^{\dag} \Sigma \mathcal{Y}_d
  q_R \right) + \text{h.c} .  \label{eq:minimal-fermion-mass}
\end{eqnarray}
The Dirac mass terms for leptons can be written in full analogy with the quark
mass term (\ref{eq:minimal-fermion-mass}), yielding neutrino Dirac mass terms of comparable magnitude.

As mentioned in Section \ref{pow-count}, consistency of the low-energy power
counting for a fermion propagator inside loops requires fermion masses to
count as $\mathcal{O} \left( p^1 \right)$ or smaller. This is possible here, thanks to the occurence of spurions in the fermion mass terms: this suggests a relation between spurion and momentum
expansion, specified by the counting rule
\begin{eqnarray}
  \xi \eta & = & \frac{m_t}{\Lambda_{\text{quarks}}} \hspace{1em} =
  \hspace{1em} \mathcal{O} \left( p^1 \right) .  \label{corres}
\end{eqnarray}

\subsection{Neutrino Dirac and Majorana masses}

Making use of the spurion $\mathcal{Z}$, one can construct $\Delta L = 2$
operators that are $S_{\text{nat}}$-invariant. The spurion $\mathcal{Z}$
responsible for the selection of the $\mathrm{U} \left( 1 \right)_Y$ subgroup,
controls {\emdash}via the parameter $\zeta$ {\emdash} the strength of these.
This leads us to the assume $\zeta \ll \xi, \eta \ll 1$.

Majorana masses of left- and right-handed neutrinos can thus be suppressed in
the present LEET, since they involve a coefficient $\zeta^2$. To obtain
left-handed neutrinos lighter than the charged fermions, one can forbid
the neutrino Dirac masses mentioned after equation (\ref{eq:minimal-fermion-mass}): this is done imposing a $\mathbbm{Z}_2$ symmetry which,
in the standard gauge simply reduces to $\nu_R \longmapsto - \nu_R$.

At this point, it is worth stressing that the {\tmem{number}} of spurions to
be introduced is entirely fixed once we have identified the higher
$S_{\text{nat}}$ symmetry, and once we ask to recover the electroweak group
$S_{\text{red}}$ by imposing constraints. In the right-handed sector, there
would a priori be various possibilities for the introduction of the expansion
parameters: the physical requirement that $B - L$ breaking effects be small
leaves us with three inequivalent possibilities for the ($B - L$)-breaking
building block $\mathcal{Z}$. This distinction in turn implies different
estimates for the $\nu_R$ masses, and therefore different cosmological
consequences {\cite{Hirn:2005fr}}.

\subsection{Comparing vertex and oblique corrections}

The vertex corrections as written in Section \ref{unwanted} would not be
invariant under $S_{\text{nat}}$. On the other hand, inserting appropriate
powers of spurions, we find the following $S_{\text{nat}}$-invariants, of
order respectively $\mathcal{O} \left( p^2 \xi^2 \right)$ and $\mathcal{O}
\left( p^2 \eta^2 \right)$
\begin{eqnarray}
  \mathi \overline{\chi_L} \gamma^{\mu} \mathcal{X}^{\dag}  \left( \Sigma
  D_{\mu} \Sigma^{\dag} \right) \mathcal{X} \chi_L, &  & \mathi
  \overline{\chi_R} \gamma^{\mu} \mathcal{Y}_{u,d}^{\dag} \Sigma^{\dag}  \left(
  D_{\mu} \Sigma \right) \mathcal{Y}_{u,d} \chi_R . 
\end{eqnarray}
This suggests that, in the Higgs-less LEET, certain vertex corrections could
be more important than oblique ones, which involve more powers of momenta or
spurions {\cite{Hirn:2005fr}}. A parametrization of vertex corrections at NLO,
using some simple assumptions about the flavor structure is presented in
{\cite{Hirn:2005fr}}: this opens the possibility of looking for non-oblique
deviations of the SM.

\section{Conclusions}

We have constructed a systematic LEET formalism for EWSB without a physical Higgs, which can be renormalized order by order in a momentum expansion. The leading order does not display deviations
from the SM. This requires a hidden symmetry $S_{\text{nat}}
\supset \tmop{SU} \left( 2 \right)_L \times \mathrm{U} \left( 1 \right)_Y =
S_{\text{red}}$, reduced to the electroweak group $S_{\text{red}}$ via
constraints. Implementing the constraints in a covariant manner requires
spurions: these can be used to introduce the small expansion parameters describing effects beyond the SM.

The spurions live in the coset space $S_{\text{nat}} / S_{\text{red}}$. The
constraint implies that spurions do not propagate and do not generate mass
terms for vector fields either, in contrast to GBs. There exists a ``standard
gauge'' in which spurions reduce to a set of constants. In the actual case of
the group $S_{\text{nat}}$, the spurions reduce to three constants, denoted
$\xi, \eta$ and $\zeta$. This reflects the structure of the coset space which,
in this case, is a product of three $\tmop{SU} \left( 2 \right)$ groups.

The $S_{\text{nat}}$-invariant constraints eliminate the nine redundant
fields, reduce the {\tmem{linear symmetry}} $S_{\text{nat}}$ to its
electroweak subgroup $S_{\text{red}}$, and induce couplings between the
symmetry-breaking and gauge/fermion sectors. The $W^{\pm}$ and $Z^0$ become
massive, whereas all fermions remain massless. In this way one recovers all
Higgs-less vertices of the SM. The main effect of $S_{\text{nat}}$ is the
elimination of all non-standard $\mathcal{O} \left( p^2 \right)$ vertices.

The expansion parameters $\xi, \eta$ and $\zeta$  play a role similar to quark masses in $\chi$PT. The
complete LEET invariant under $S_{\text{nat}}$ is defined as a double
expansion: in powers of momenta and in powers of spurions. The LEET at leading
order coincides with the Higgs-less vertices of the SM, used at tree level.
Majorana and Dirac mass terms, which could appear at $\mathcal{O} \left( p^1
\right)$, now involve in addition powers of
spurions: within the expansion, these operators can be consistently counted as
$\mathcal{O} \left( p^2 \right)$ or higher. Also, non-standard $\mathcal{O}
\left( p^2 \right)$ vertices reappear as $S_{\text{nat}}$-invariant operators
explicitly containing spurions, i.e. suppressed by powers of the parameters
$\xi, \eta$ and $\zeta$ in addition to $p^2$. In particular, we note that
vertex corrections appear before oblique ones (the latter cannot be
disentangled from loops).

The consequences of the assumed higher symmetry can be studied systematically using  the formalism proposed here for the Higgs-less case: vertex corrections,  lepton-number violation  as well as the cosmological consequences of light right-handed neutrinos \cite{Hirn:2005fr}.

{\section*{Acknowledgments}}

We acknowledge support by the European Union EURIDICE network (HPRN-CT-2002-00311).\\

{

\end{document}